\documentstyle[11pt]{article}
\def\beq{\begin{eqnarray}}
\def\eeq{\end{eqnarray}}
\def\bea{\begin{eqnarray*}}
\def\eea{\end{eqnarray*}}

\def\centeron#1#2{{\setbox0=\hbox{#1}\setbox1=\hbox{#2}\ifdim
\wd1>\wd0\kern.5\wd1\kern-.5\wd0\fi
\copy0\kern-.5\wd0\kern-.5\wd1\copy1\ifdim\wd0>\wd1
\kern.5\wd0\kern-.5\wd1\fi}}
\def\ltap{\;\centeron{\raise.35ex\hbox{$<$}}{\lower.65ex\hbox{$\sim$}}\;}
\def\gtap{\;\centeron{\raise.35ex\hbox{$>$}}{\lower.65ex\hbox{$\sim$}}\;}

\newcommand{\newc}{\newcommand}
\newc{\qbar}{{\overline q}}
\newc{\Kahler}{K\"ahler }
\newc{\deltaGS}{\delta_{\rm GS}}
\begin{document}
\begin{titlepage}
\begin{flushright}
{\large hep-th/yymmnnn \\ SCIPP-2009/01\\
}
\end{flushright}

\vskip 1.2cm

\begin{center}

{\LARGE\bf 
Supersymmetry Breaking at Low Energies}

\vskip 1.4cm

{\large Michael Dine}
\\
\vskip 0.4cm
{\it Santa Cruz Institute for Particle Physics and
\\ Department of Physics, University of California,
     Santa Cruz CA 95064  } \\
\vskip 4pt

\vskip 1.5cm

\begin{abstract}
These lectures, given at the Cargese Summer school
in 2008,  provide an introduction to dynamical supersymmetry
breaking and gauge mediation, with emphasis on the recent appreciation
of the possible role of metastable supersymmetry breaking, and
the evolving understanding of General Gauge Mediation.  The underlying focus is on how supersymmetry might be realized at the Large Hadron Collider.
\vspace{1pc}
\end{abstract}


\end{center}

\vskip 1.0 cm

\end{titlepage}
\setcounter{footnote}{0} \setcounter{page}{2}
\setcounter{section}{0} \setcounter{subsection}{0}
\setcounter{subsubsection}{0}

\section{Supersymmetry on the Eve of the LHC}

As we await collisions at the LHC, it seems a good time to assess various proposals for physics beyond the Standard Model.  Certainly the possibility which has gained
the most attention is that nature is approximately supersymmetric, with supersymmetry broken at the TeV scale.  There are at least four reasons for this:
\begin{enumerate}
\item  Supersymmetry can naturally account for the enormous hierarchy between the Planck scale and the weak scale.
\item  With the assumption that all new thresholds lie at about $1$ TeV, the gauge couplings unify reasonably well, at a scale of order $10^{16}$ GeV.
\item  With the additional assumption of a conserved $R$ parity (the simplest hypothesis through which to forbid rapid proton decay), the theory
automatically possesses a dark matter candidate, produced in abundance comparable to the observed dark matter density.
\item  Supersymmetry arises rather naturally in many string constructions, and might plausible be broken by low energy, non-perturbative
dynamics.
\end{enumerate}

Each of these points, however, is open to serious challenge, and one is entitled to be -- indeed should be -- skeptical that we are on the brink
of the discovery of an extraordinary, previously unknown, symmetry of nature.
\begin{enumerate}
\item  Current experimental constraints still require some fine tuning of Higgs parameters -- as severe as $1\%$ in many models.  This is referred
to as the ``little hierarchy" problem.
\item  Unification is not quite perfect, and depends on details of threshold effects at high and low scales.  Unification itself, within, say our current understanding
of string theory, is not automatic and it is unclear in what sense it is generic.   Still, the success of the unification calculation strikes many physicists as remarkable.
\item  Some degree of tuning is often required, given experimental constraints, to actually obtain the observed dark matter density.
\item   As string theorists have explored the so-called string ``landscape", questions have been raised as to just how generic low energy supersymmetry
may be in string theory.  They have also raised the worry that the hierarchy problem may ultimately have an anthropic explanation, similar to that
which has been offered for the cosmological constant\cite{weinbergcc}.
\end{enumerate}

While theorists debate the significance of each of these points, there is little dispute that
the various alternative proposals (technicolor, large or warped extra dimensions) have even more
serious problems.   In these lectures, we will adopt the optimistic viewpoint
that these challenges may provide hints as to how supersymmetry is realized in nature.
As we will review, the last few years have seen significant progress in understanding
supersymmetric dynamics, and have opened up new possibilities for model building.  So it is
a good time to attempt to construct complete theories of lo energy supersymmetry and its breaking.

\section{Challenges for Supersymmetry}

If the challenges to supersymmetry are clues to the underlying structure, it is important to understand them well.  This is the focus of this section.

\subsection{The Little Hierarchy}

The little hierarchy may well indicate that theorists are on the wrong track.  But if the supersymmetry hypothesis
is correct, it suggests, we will see, a relatively low scale of supersymmetry breaking.
The issues are nicely illustrated by ignoring gauge interactions, and just examining the top quark Yukawa couplings.  From the top quark loop,
one gets a quadratically divergent correction to the Higgs mass:
\beq
\delta m_{H_U}^{2} = -6\vert y_t \vert^2 \int^\Lambda {d^4 k \over (2 \pi)^4} {1 \over k^2-m_t^2},
\eeq
where $\Lambda$ is some cutoff.
Without supersymmetry, This result is proportional to some cutoff scale squared.  This is the usual statement of the hierarchy problem;  without supersymmetry,
masses of scalars are expected, by dimensional analysis, to be of order $\Lambda^2$, where $\Lambda$ is some scale where the effective lagrangian
which describes their interactions breaks down (e.g. due to compositeness, some modification of the structure of space-time, or something equally dramatic).

With supersymmetry,
the diagrams with the stops give a contribution which cancels the leading divergence:
\beq
\delta m_{H_U}^{2{(2)}} = 6 \vert y_t \vert^2 \int {d^4 k \over (2 \pi)^4}{1 \over k^2 - \tilde m_t^2}.
\eeq
There is still a subleading logarithmic divergence, which, for $\tilde m_t^2 \gg m_t^2$ yields:
\beq
\delta m_{H_U}^2 \approx  -  6 \tilde m_t^2 {y_t^2 \over 16 \pi^2} \ln({\Lambda^2 \over \tilde m_t^2})
\label{littlehierarchy}
\eeq
where $\Lambda$ is an ultraviolet cutoff, and we have neglected $m_t^2$ relative to $\tilde m_t^2$.
The quadratic divergence has been replaced by a logarithmic one;  the severe hierarchy problem which exists in the absence
of supersymmetry is solved, provided that the scale of supersymmetry breaking -- the masses of the
various superpartners -- is not much greater than the weak scale.

Without supersymmetry, assuming the cutoff is large, say of order $M_{gut} = 2 \times 10^{16}$ GeV, the correction to the Higgs mass is enormous; in order that the
mass be of order, say, $100$ GeV, one needs to introduce a bare mass parameter of order $10^{30}$ GeV$^2$, which cancels against the radiative correction up to
one part in $10^{26}$.  So supersymmetry yields an enormous improvement.
But there is still a difficulty.   As we will explain shortly, within simple supersymmetric models, the physical
Higgs mass cannot be much larger $m_Z$.  Let's examine eqn.
\ref{littlehierarchy} in view of this constraint.  Limits on the stop mass are in the $350$ GeV
range.  If we take the cutoff, $\Lambda$, to be of order $M_0$, we then obtain a correction
\beq
{\Delta m_{H_U}^2 \over m_Z^2} \approx 40
\eeq
i.e. a tuning of order 2\%.  But the situation may be even worse than this; in many models,
there are arguments that the stop mass should be significantly larger, $800$ GeV or more.  In that case, the fine tuning
is more like a part in $200$!

If we are to ameliorate this, we need to suppress both the logarithm and the stop mass.  For example, suppose that the cutoff is $10$ TeV (we will motivate
this choice shortly), and that the stop mass is close to the experimental limit.  Then $\Delta m_{H_U}^2 \sim 4 m_Z^2$,
arguably not fine tuned at all.

What, physically, is the cutoff scale $\Lambda$?   In so-called supergravity models, as they are not renormalizable, it is naturally thought of as of order the Planck
scale (perhaps the string scale, if we imagine the theory is embedded in string theory).  As we will see, however, the cutoff can be much lower.  Indeed, in general, the
scale is related to the scale associated with the {\it messengers} of supersymmetry breaking.

These considerations provide one motivation for thinking about supersymmetry breaking at low scales.  Indeed, they suggest that
supersymmetry should be broken at an underlying scale close to the $100$'s of GeV  scale of squarks and sleptons.
A widely studied framework which might achieve this is known as {\it gauge mediation}\cite{gaugemediation}.

\subsection{The Proliferation of Parameters}

My experimental colleagues often express frustration with theorists:  they have already presented us with $20$ or so parameters to explain, and rather
than do so, we seem committed to introducing more.  At a minimum,  supersymmetry introduces $105$ new parameters.
Counting of these parameters in the simplest generalization of the Standard Model, the {\it Minimal Supersymmetric
Standard Model} (MSSM) is very simple, and worth doing.   I'll define the MSSM by its particle content:  it is a theory in which each fermion of the standard
model is replaced by a chiral superfield, each gauge boson by a vector field, and the higgs bosons by a pair of chiral fields, $H_U,~H_D$.  The gauge quantum numbers of
the quark and lepton superfields are just those of the Standard Model; the Higgs superfields, $H_U$ and $H_D$, are doublets, with hypercharge $\pm 1$.  (See, for example,
\cite{dinebook}, chapter 11,  for an introduction).
It is necessary to have at least two Higgs doublets in order to avoid anomalies (perturbative and non-perturbative).  In order to explain the observed
features of quark and lepton masses, it is natural to suppose that the superpotential contains a generalization of the Standard Model Yukawa couplings:
\beq
W_{y} = y_U H_U Q \bar U + y_D H_D Q \bar D + y_L H_D \bar E .
\eeq
Here $y_U$ and $y_D$ are $3 \times 3$ matrices in the space of flavors.

But additional couplings are permitted, and here we encounter some troubling features.  First, for the Higgs fields, we can, and will, add a ``$\mu$" term,
\beq
W_\mu = \mu H_U H_D.
\eeq
If $\mu$ is too large, the Higgs fields are too massive to play a role in electroweak symmetry breaking; if it is too small, one can't
obtain an
acceptable chargino and neutralino spectrum.  In supersymmetric theories, as I'll explain, there is at most logarithmic renormalization of $\mu$, but still, dimensional
analysis would suggest that it should be of
order GUT or  Planck scale.  One might imagine it's small value arises due to an underlying symmetry; string theory has also provided
a variety of mechanisms which might provide an explanation.  Still, the fact that this is not automatic is, at first sight, disappointing.

Second, there are a class of couplings which violate lepton and/or baryon number:
\beq
W_{b/l-violating} = \Gamma_1 \bar U \bar D \bar D + \Gamma_2 Q L \bar D  + \Gamma_3 L L \bar E.
\eeq
The couplings $\Gamma$ are dimensionless, and thus might be expected to be of order one -- and hence catastrophic.  The simplest
approach to this problem is to banish these couplings entirely:  impose a discrete symmetry, called ``R parity", under which the partners
of ordinary fields (squarks, sleptons, gauginos, higgsinos) are odd, and ordinary fields are even.  Again, it is somewhat disappointing that
one needs to impose yet another requirement on these theories, but at least symmetries are natural, and indeed symmetries of this type
often arise in string theory.  Moreover, imposing this symmetry brings a bonus:  the lightest of the new supersymmetric particles, the LSP,
is stable, and potentially a candidate for the dark matter.  It should be noted, as well, that other discrete symmetries which treat fermions
and bosons differently ($R$ symmetries) can achieve this result, with other potentially interesting consequences (e.g. such a symmetry
might explain the smallness of $\mu$, or the very strong limits on the proton lifetime).

With these caveats, we are ready to count parameters.  So far, we have roughly the same number of parameters as the Standard Model.
But supersymmetry is clearly broken in nature, and the masses of squarks, sleptons and gauginos are subject to severe experimental constraints.
So we add to the lagrangian a set of soft supersymmetry breaking terms.  The allowed terms (those which don't reintroduce power law divergences
are easily classified\cite{grisarusoft,dimopoulosgeorgi}; see \cite{dinebook}, pp. 162-163 for a simple derivation), and include:
\begin{enumerate}
\item  Soft mass terms for squarks, sleptons, and Higgs fields:
\beq
{\cal L}_{scalars}= Q^* m_Q^2 Q + \bar U^* m_U^2 \bar U + \bar D^* m_D^2 \bar D
\eeq
$$~~~~+ L^* m_L^2 L
 + \bar E^* m_E \bar E $$
 $$~~~~
+ m_{H_U}^2 \vert H_U\vert^2 + m_{H_U}^2 \vert H_U\vert^2 + B_\mu H_U H_D + {\rm c.c.}
$$
$m_Q^2$, $m_U^2$, etc., are matrices in the space of flavors.
The first five matrices are $3\times 3$ Hermitian matrices ($45$ parameters); the Higgs mass terms add an additional $4$, for a total of $49$ parameters.
\item  Cubic couplings of the scalars:
\beq
{\cal L_A} = H_U Q ~A_U ~\bar U +H_D Q ~A_D ~\bar D
\eeq
$$~~~~ +H_D L ~A_E
 \bar E + {\rm c.c.}
$$
Here the matrices $A_U$, $A_D$, $A_E$ are complex matrices, so we have an additional $54$ parameters.
\item  Mass terms for the U(1) ($b$), $SU(2)$ ($w$), and $SU(3)$ ($\lambda$) gauginos:
\beq
m_1 b b + m_2 w w + m_3 \lambda \lambda
\eeq
These are three complex quantities, making six additional parameters.
\end{enumerate}
So we would seem to have an additional $109$ parameters.  However, the supersymmetric part of the MSSM
lagrangian has symmetries which are broken by the general soft breaking terms (including $\mu$ among the
soft breakings):
\begin{enumerate}
\item  Two of three separate lepton numbers
\item  A ``Peccei-Quinn" symmetry, under which $H_U$ and $H_D$ rotate by the same phase, and the quarks and leptons transform suitably.
\item  A continuous "$R$" symmetry, which we will explain in more detail below.
\end{enumerate}
Redefining fields using these four transformations reduces the number of parameters to $105$.

\subsection{Aside on R Symmetries}

In the previous section, we mentioned the $R$ symmetry of the MSSM which exists in the absence of soft breakings.  In general, an $R$ symmetry
is a symmetry under which the supercharges rotate (equivalently, for which the generator does not commute with the supercharges).  This is possible
since the Hamiltonian is quadratic in the charges.  Under a continuous $R$ symmetry, if we define the Grassmann coordinates, $\theta$, to transform with
phase $e^{i\alpha}$, the supercharges $Q_\alpha$ transform with the opposite phase, and the superpotential transforms with phase $e^{2i\alpha}$.
A model respects a continuous $R$ symmetry if it is possible to assign charges to all fields so that the Kahler potential is invariant and the superpotential transforms with
$R$ charge $2$.  Gauginos transform with charge $1$; the fermions in chiral multiplets transform with one unit less charge than the scalars.

In the case of the supersymmetric couplings of the MSSM, including the $\mu$ term, one can assign:  $H_U,H_D$ $R$ charge $1$ and all of the quark
and lepton superfields $R$ charge $1/2$.  (Other assignments differ by combinations of baryon number and other symmetries).

In general, we don't expect global continuous symmetries in fundamental theories; when they arise, as in the case of baryon number in the Standard Model,
they should be accidental consequences of renormalizability and other symmetries.  On the other hand, {\it discrete} R symmetries are plausible; they are quite
common in string theory, for example (arising, for example, as unbroken subgroups of higher dimensional rotation groups upon compactification).  At the
level of renormalizable couplings, such symmetries could well lead to continuous symmetries.  We have now seen that this occurs in the MSSM, once one
supposes an $R$ parity (perhaps the simplest of discrete $R$ symmetries).

\subsection{Constraints on the Soft Parameters}

Ignoring the Higgs Yukawa couplings, the fields of the Standard Model respect a large set of flavor symmetries.
Since the quark masses (apart from the top quark) are small compared
to the masses of the $W$ and $Z$ bosons, these flavor symmetries play an important role in the phenomena of weak
interactions.  A priori, there is no reason for the squark and slepton masses to respect these symmetries.  But if these masses
are of order the weak scale, and are simply random numbers, the effects on flavor changing processes can
be dramatic.

The most stringent constraints come from study of the neutral kaon system.  One of the early triumphs of the Standard Model was its explanation
of the small rates of ``neutral current" processes in this system ($K^0 \leftrightarrow \bar K^0$, $K^0 \rightarrow \mu \bar \mu$, etc.).
These effects are suppressed not only by $G_F$ and a loop factor, but, due to the approximate
flavor symmetries mentioned above, an additional factor of $m_c^2/M_W^2$, i.e. nearly $10^{-4}$.
Supersymmetry introduces many new contributions to these processes.  Exchange of gluinos and squarks, for example, gives a contribution
to $K$-$\bar K$ mixing which is suppressed, for general values of the masses, only by
${\alpha_s \over 4 \pi}{1 \over m_{susy}^2}$, not nearly enough.

\vskip .2cm
\noindent
{\bf  Exercise:}  Estimate the gluino box contribution to $K \bar K$ mixing by comparing to the ordinary weak interactions.  Unless you are careful
about chirality, you will underestimate the contribution by nearly an order of magnitude!

\vskip .3cm

These problems are further exacerbated when one allows for CP violating phases in the soft terms.  If these phases are simply numbers of order one,
the constraints become tighter by an order of magnitude.

One can satisfy all of these constraints at once if one assumes that the squark and slepton masses are all degenerate at some energy scale.
More precisely, one assumes that each of the matrices
$m_Q, m_{\bar U}$, etc., are proportional to the unit matrix, while the matrices $A_U,A_D$, etc., are proportional to the Yukawa couplings.
This automatically suppresses flavor changing processes in the light meson systems; any violation of flavor is proportional to small Yukawa
couplings and mixing angles.    Masiero\cite{masiero} carefully reviews the constraints on these processes, and translates them into
constraints on the degree of degeneracy of squark and slepton masses.

So if supersymmetry breaking is at a low scale, the $105$ or more new parameters associated with supersymmetry are highly constrained.
Most analyses of supersymmetry phenomenology and model building simply take some level of degeneracy as a given.  For example,
most experimental studies are based on the assumption that the spectrum can be described by three parameters,
$m_0^2, m_{1/2}, A,\mu$, the first being a common value of the scalar masses at some high scale (e.g. the unification scale), the second
the common value of the gaugino masses, the third a common $A$ term, and the fourth the $\mu$ term in the superpotential we have discussed earlier.

One can ask how natural is this assumption?  The set of parameters described above are often called "minimal supergravity", the idea being that
gravity is universal, and supergravity should be as well.  But, as we will see in section
\ref{supergravity}, a supergravity theory with the field content at low energies of the MSSM has precisely
the number of independent parameters we have counted above, and there is no symmetry which forces any degree of degeneracy.
Still, except for the top Yukawa, all of the Yukawa couplings are small, and it seems plausible that there might be some approximate flavor symmetry
which could account for degeneracy.  Microscopic models with such symmetries have been considered by many authors, and dynamical explanations
have been suggested as well for accidental
flavor symmetries (particularly involving large extra dimensions).  But probably the simplest implementation is provided by models
of ``gauge mediation"\cite{gaugemediation}.  Gauge mediated models can be defined as field theory models where, in the limit that the gauge couplings
tend to zero, there is no coupling between the dynamics which breaks supersymmetry (``hidden sector") and the fields of ordinary physics\cite{meadeseibergshih}.
Given the assumption that there is a good effective field theory description at a scale below that which determines the quark and lepton Yukawa
couplings, these couplings {\it can} be treated as perturbations.  This definition, as we will discuss, requires modification in order to account for
the $\mu$ term, and any modification risks reintroducing flavor violations.  But this simple success in providing a generic framework for the
absence of flavor changing neutral currents is one of the principle reasons for interest in gauge mediation.

That said, flavor violation in light meson systems is not the only constraint which low energy physics places on the soft breaking parameters.
There are at least two additional striking facts which any model must account for.  First are the constraints on electric dipole moments of quarks (the neutron)
and the electron.   For example, for quark electric dipole moments one might expect
\beq
d_q \approx {\alpha_s \over 4 \pi} {m_q \over m_{\lambda}^2} \sin(\delta)
\eeq
where the gluino mass here represents some average of masses appearing in the one loop contribution, and $\delta$ is a CP-violating phase.  For the $d$ quark,
and assuming a large value, say $1$ TeV for the gluino mass, one obtains about $2 \times 10^{-24} \sin \delta$ e-cm, which is more than
an order of magnitude above the experimental
limit; things are significantly worse if the gluino is lighter.  So one needs to somehow explain the suppression of the phase $\delta$.
The second constraint is the rate for the inclusive process $b \rightarrow s + \gamma$.  Current experimental data and Standard Model
calculations leave a small amount of room for possible new physics contributions.  However, the charged Higgs contribution in the MSSM tends
to be too large, unless the Higgs is rather heavy\cite{bsgamma}.

\subsection{The Higgs Mass}

Since the conclusion of the LEP II program, there is a very severe constraint on models of supersymmetry:  the Higgs mass itself.  In the MSSM,
as we have remarked,
it is easy to show that there is a strict limit on the Higgs mass at the classical level:  $m_H < M_Z$.  This result is not surprising.
In the Standard Model, the Higgs mass is proportional to the Higgs quartic coupling; in the MSSM,
  the Higgs quartic couplings arise entirely from gauge interactions.  Because the top Yukawa coupling is so
large (of order one), however, there is the potential for large radiative corrections.  One loop corrections generate a
substantial supersymmetry-violating Higgs
quartic coupling.  It is easy to isolate
a contribution  which diverges logarithmically as the stop mass tends to infinity.  In the large
$\tan \beta$ limit, and assuming that the left and right-handed stop quark masses are identical, there is a correction to the $\vert H_U \vert^4$
coupling:
\beq
\delta \lambda = {3 y_t^4 \over 16 \pi^2} \ln (\tilde m_t^2/m_t^2).
\label{deltalambda}
\eeq

\vskip .2cm
\noindent
  {\bf Exercise:}  Verify eqn. \ref{deltalambda}.
\vskip .2cm

Plugging this into the potential, and working out the corresponding physical Higgs mass, one finds that requiring a Higgs mass of at least $116$ GeV
requires that $\tilde m_t > 800$ GeV.  From the point of view of fine tuning we described earlier, this is troubling.  More complete computations
are readily found in the literature.  If one allows significant mixing between the left and right-handed stops, the situation is not necessarily as severe
(for a recent discussion, see \cite{essig}).

Alternatively, additional fields and interactions can alter the situation.  The simplest possibility, known as the NMSSM (for Next to Minimal Supersymmetric
Supersymmetric Standard Model) contains an additional singlet, $S$, with coupling $\lambda_S S H_U H_D$.  This provides an additional quartic Higgs
couplings, and can enlarge (slightly) the allowed range of Higgs masses.

More generally, if there is some new physics at a scale, $M$, slightly above the scale of the Higgs fields, one can integrate out this physics and represent its
effects in a local effective lagrangian\cite{dst}.  Corrections to quartic couplings will be down by powers of $M$, so the lowest dimension operators will be most important.
There is a unique operator of dimension five which generates a quartic coupling:
\beq
{\cal O} = {1 \over M} \int d^2 \theta H_U H_D H_U H_D.
\eeq
Solving for the auxiliary fields, gives a quartic coupling, of order $\mu/M$.  The NMSSM can be understood in this way, in the case that the mass of the singlet is
larger than $\mu$.  These terms are most important when $H_U \sim H_D$ (the region of ``small $\tan \beta$").  If $H_U \gg H_D$, dimension
six operators are most important.  In either case, such effects can yield a Higgs of mass greater than the experimental bound, but masses much larger
than that seem hard to achieve.

\section{Supergravity Models}
\label{supergravity}

Despite our somewhat disparaging remarks about supergravity models and the problem of flavor, these theories provide a simple conceptual
framework in which to consider supersymmetry breaking and its mediation to low energy physics.  They are a natural setting, as well, to think
about the possible phenomenologies which might emerge from string theory.  The bosonic part of the general supergravity lagrangian, including
terms with at most two derivatives, can be described rather simply.  The lagrangian is specified by three functions: the Kahler potential,
superpotential and gauge coupling function, $K(\Phi_i, \Phi_i^*)$, $W(\Phi_i)$, and $f_a(\Phi_i)$.  The Kahler potential is a general function
of the fields, while the superpotential and gauge coupling function (the latter is really several functions) are holomorphic.  The scalar potential
is given by:
\beq
V= e^{K/M_p^2}  [ \left ( {\partial W \over \partial \Phi_i} + {1 \over M_p^2}{\partial K \over \partial \Phi_i} W \right ) g^{i \bar j}
\eeq
$$~~~~~~
\left ( {\partial W \over \partial \Phi_{\bar j}^*} + {1 \over M_p^2}{\partial K \over \partial \Phi_{\bar j}^*} W^* \right ) - 3 \vert W \vert^2 ]
$$
to which must be added the terms proportional to $(D^a)^2$ from the gauge couplings.
 The condition for unbroken susy is that the ``Kahler derivative" of $W$ with respect to all fields should vanish:
 \beq
 D_i W = 0~~~~D_i W = {\partial W \over \partial \phi_i} + {\partial K \over \partial \Phi_i} W.
\eeq
In addition, the auxiliary fields for all of the gauge groups must vanish.

Now to make a model of low energy physics, we assume that
\begin{enumerate}
\item Supersymmetry is broken in a hidden sector, involving some fields $X_\alpha$; the visible sector fields will be denoted $\phi_i$.
\item The Kahler potential has a very simple form:
\beq
K = \sum \phi_i^* \phi_i + X_\alpha^* X_\alpha
\eeq
This potential has a $U(N)$ symmetry, where $N$ is the number of fields; this symmetry will suppress flavor changing neutral currents.
The scalar mass terms are universal, and the soft breaking cubic terms are proportional to the fermion Yukawa couplings.
\item  The superpotential is a sum of hidden sector and visible sector fields:
\beq
W = f(\phi_i) + g(X_\alpha).
\eeq
\item
 $g(X_\alpha)$ is assumed to break supersymmetry, $D_\alpha W \ne 0$.
 \item
 Gaugino masses arise from a universal gauge coupling function of the form
 \beq
 f(X) = f(0) + c_\alpha {X_\alpha \over M}.
 \eeq
\item  A $\mu$ term can arise from a term in the Kahler potential such as $X^\dagger H_U H_D$.  Such couplings are often generated
by integrating out heavy fields.
 \end{enumerate}

 Before listing the problems of these theories, some of their features should be noted.  The mass of the gravitino, assuming that the space-time is flat,
 is given by
 \beq
 m_{3/2} = \langle e^{K/2} W_0 \rangle.
 \eeq
 Scalar masses arise from terms in the potential such as
 \beq
 \left \vert {\partial W \over \partial \phi_i} + \phi_i^* W \right \vert^2 \sim \vert \phi_i  \vert^2 m_{3/2}^2.
 \eeq
 $A$ terms arise from a variety of sources, such as
 \beq
\left ( { \partial W \over \partial X} \right )^*{ \partial K \over \partial X} W
\eeq
where one takes the term in $W$ which is cubic in fields ($H_U Q \bar U$, etc), and replacing $X$ by its vev.

The assumptions underlying the model, however, are quite strong.
As we have remarked, the $U(N)$ symmetry assumed in the Kahler potential cannot reasonably be expected to hold in a general situation,
even approximately.  Terms like $X X^\dagger Q Q^\dagger$, with various flavor indices, will generate new, non-universal
terms in the mass matrices and $A$ terms.
In string theory, various scenarios have been put forth which might give rise to accidental symmetries of the type required,
but it is not at all clear how plausible these are.

\section{Minimal Gauge Mediation}

We have already described the main premiss underlying gauge mediation:  in the limit that the gauge couplings vanish, the hidden and visible sectors decouple.
A simple model illustrates the basic idea.  Suppose we have a chiral field, $X$, with
\beq
\langle X \rangle = x + \theta^2 F.
 \eeq
 Suppose also that $X$ is coupled to a vector-like set of fields, transforming as a single $5$ and $\bar 5$ of $SU(5)$:
 \beq
 W = X(\lambda_\ell \bar \ell \ell + \lambda_q \bar q q).
\label{simplegm}
\eeq
For $F<X$, $\ell, \bar \ell, q, \bar q$ are massive, with supersymmetry breaking splittings of order $F$.
The fermion masses are given by:
\beq
m_q = \lambda_q x~~~ m_\ell = \ \lambda_\ell  x
\eeq
while the scalar splittings are
\beq
\Delta m_q^2 = \lambda_q F ~~~~~ \Delta m_\ell^2 = \lambda_\ell F.
\eeq

 In such a model, masses for gauginos are generated at one loop; for scalars at two loops.  The gaugino mass computation
 is quite simple.  Even the two loop scalar
 masses turn out to be rather easy, as one is working at zero momentum.  The latter calculation
can be done quite efficiently using supergraph
techniques; an elegant alternative uses background field arguments\cite{masscalculations}.
The result for the gaugino masses is:
\beq
m_{\lambda_i} = {\alpha_i \over \pi} \Lambda,
\eeq
while for the squark and slepton masses it is:
\beq
\widetilde m^2 ={2 \Lambda^2}
[
C_3\left({\alpha_3 \over 4 \pi}\right)^2
+C_2\left({\alpha_2\over 4 \pi}\right)^2
\label{scalarsmgm}
\eeq
$$~~~
+{5 \over 3}{\left(Y\over2\right)^2}
\left({\alpha_1\over 4 \pi}\right)^2 ],
$$
where $\Lambda = F_x/x$.
$C_3 = 4/3$ for color triplets and zero for singlets,
$C_2= 3/4$ for
weak doublets and zero for singlets.

This spectrum has a number of notable features.
\begin{enumerate}
\item  One parameter describes the masses of the three gauginos and the squarks and sleptons
\item  Flavor-changing neutral currents are automatically suppressed; each of the matrices $m_Q^2$, etc., is automatically proportional to the
unit matrix; the $A$ terms are highly suppressed (they receive no one contributions before three loop order).
\item  CP conservation is automatic
\item  This model cannot generate a $\mu$ term; the term is protected by symmetries.  Some further structure is necessary.
\end{enumerate}

\subsection{Messenger Parity and $D$ terms}

The model of eqn. \ref{simplegm} neatly avoids a potential pitfall of this sort of construction.  There are, in fact, individual one loop
graphs which contribute to squark and slepton masses, corresponding to the possibility of an expectation value for
the auxiliary field of the hypercharge multiplet, $\langle D \rangle$.  $\langle D \rangle$ receives contributions from
messenger fields, and can be non-vanishing because supersymmetry is broken.  However, this model
possesses an accidental, approximate symmetry under which
\beq
q \leftrightarrow \bar q ~~~ \ell \leftrightarrow \bar \ell ~~~~ V_Y \rightarrow -V_Y
\eeq
This symmetry is broken by the interactions of the MSSM, but this will only be visible at high loop order.
It is easy to check the cancelation of the corresponding diagrams explicitly.

\subsection{Minimal Gauge Mediation and the Little Hierarchy Problem}

In considering the little hierarchy, the first question to ask is ``what is the cutoff $\Lambda$ in eqn. \ref{littlehierarchy}.
In renormalizable theories like those considered here, there cannot actually be a divergence.  The one loop correction
to the Higgs mass is really a three loop graph, in which the two loop subgraph responsible for the stop mass has been
shrunk to a point.  This description breaks down at the scale $M$ (the scale of the messenger masses).  So, in MGM,
$M$ is the cutoff.  $M$, can in principle be as small as $M \sim {4 \pi \over \alpha_s} ({\rm few~hundred~GeV})
\sim {\rm few~tens~of~TeV}$.  So the log need not be terribly large.

On the other hand, given the experimental constraints on the susy spectrum, the stop mass appearing in the loop is necessarily
quite large in MGM.
In particular, the experimental limits on the mass of the lightest charged slepton are of order $100$ GeV (the precise limit depends on
the allowed decay channels).
Satisfying this constraint implies, from the formulae for the scalar masses, that the squark masses are quite large, of order
$800$ GeV.  So one has a fine tuning, at best, at the few per cent level.

If one could {\it compress} the spectrum, one could ameliorate the little hierarchy.  In other words, if one had a spectrum in which
squark and slepton masses were not so different, one could improve the situation significantly.  Current experimental limits
on the squark masses are $300-400$ GeV, so fine tuning of less than a part in five may be possible.  We will see
shortly that such a compression of the spectrum can occur in more general models of messengers.

\section{General Gauge Mediation}

Much work has been devoted to understanding the properties of this simple model, but it is natural to ask:
just how general are these features?  It turns out that they are peculiar to our assumption of a single set of messengers
and just one singlet responsible for supersymmetry breaking and R symmetry breaking.
Meade, Seiberg and Shih\cite{meadeseibergshih} have formulated the problem of gauge mediation in a general way,
and dubbed this formulation {\it General Gauge Mediation}  (GGM).    They study the problem
in terms of correlation functions of (gauge) supercurrents.  Analyzing the restrictions imposed by Lorentz invariance and supersymmetry
on these correlation functions, they find that the general gauge-mediated spectrum is described by three complex parameters and three real
parameters.

While we won't review the analysis of \cite{meadeseibergshih} in detail, it is easy to see, in simple weakly coupled models, how one can obtain a larger set of parameters.
Take, for example, a model, as above, with messengers $q,\bar q, \ell,\bar \ell$, but replace the one singlet of the earlier
model with a set of singlets, $X_i$.  For the superpotential, take:
\beq
W = \lambda_i^q X_i \bar q q + \lambda_i^\ell X_i \bar \ell \ell.
\eeq
Now, unlike the case of minimal gauge mediation, the ratio of the splittings in the multiplets to the average (i.e. fermion) masses is not the same
for $q,\bar q$ and $\ell, \bar \ell$.  For the fermion masses:
\beq
m_q = \sum \lambda_i^q x_i ~~~ m_\ell = \sum \lambda_i^\ell  x_i
\eeq
while the scalar splittings are
\beq
\Delta m_q^2 = \sum \lambda_i^q F_i ~~~~~ \Delta m_\ell^2 = \sum \lambda_i^\ell F_i.
\eeq
In the case of MGM, the one loop contributions for fields carrying color were proportional to $\Delta m_q^2/m_q^2$,
while those contributing to $\ell$ were proportional to $\Delta m_\ell^2/m_\ell^2$.
One now finds, simply generalizing the previous computation, for the masses of the gauginos:
\beq
m_{\lambda} = {\alpha_3 \over 4 \pi} \Lambda_q~~~~
m_w = {\alpha_2 \over 4 \pi} \Lambda_{\ell}
\label{gluinoformula}
\eeq
$$~~~~
m_b =  {\alpha_1 \over 4 \pi} \left [2/3 \Lambda_q +
\Lambda_\ell \right ].
$$
where
\beq
\Lambda_q = {\lambda_q^i F_i \over \lambda_q^j x_j}~~~
\Lambda_\ell = {\lambda_\ell^i F_i \over \lambda_\ell^j x_j}
\eeq
($i$ and $j$ summed).
Similarly, for the squark and slepton masses we have:
\beq
\widetilde m^2 =2
[
C_3\left({\alpha_3 \over 4 \pi}\right)^2 \Lambda_q^2
+C_2\left({\alpha_2\over 4 \pi}\right)^2 \Lambda_\ell^2
\label{scalarsggm1}
\eeq
$$~~~~
+{\left({Y\over2}\right)^2}
\left({\alpha_1\over 4 \pi}\right)^2 ({2 \over 3} \Lambda_q^2 + \Lambda_\ell^2) ]
$$

At this point, it is easy to understand the parameter counting of Meade et al.  We can write the general gauge-mediated spectrum in terms of three independent
complex masses for the gauginos, and parameterize
the general sfermion mass matrix as:
\beq
\widetilde m^2 =2
[
C_3\left({\alpha_3 \over 4 \pi}\right)^2 \Lambda_{qcd}^2
+C_2\left({\alpha_2\over 4 \pi}\right)^2 \Lambda_w^2
\label{scalarsggm}
\eeq
$$~~~~
+{\left({Y\over2}\right)^2}
\left({\alpha_1\over 4 \pi}\right)^2 \Lambda_b^2 ],
$$
In the present case, there are two relations among these masses, which can be expressed as sum rules.  But more generally, we have
three independent complex parameters, the gaugino masses, and three additional real parameters.

Models with additional fields permit independent values for all of the parameters of GGM.
In constructing examples, we will insist that the messengers fill complete multiplets of $SU(5)$, so as to preserve unification (one can legitimately
ask why nature would be so concerned with achieving unification).  For example, suppose one has a $10$ and $\bar 10$
of messengers, and multiple singlets.  The messengers can be denoted as $Q,\bar Q$, $U, \bar U$, and $E,\bar E$.
One now has three independent parameters, which, by analogy to our previous example, can be denoted as $\Lambda_Q,\Lambda_U,\Lambda_E$.
 The minimal, weak coupling theory which yields
 the full set of parameters of GGM consists of a $10$ and $\bar 10$, and two $5, \bar 5$ pairs.  In this case, however,
if the scale of supersymmetry breaking is low, the gauge couplings tend to get strong well below the unification scale.
In addition, there is not an automatic messenger parity, so it is necessary to require additional structure in order to suppress the
Fayet-Iliopoulos term for hypercharge.  Finally, these models don't actually cover the full parameter space (though they
have the maximal number of parameters); a strategy for doing this is described in \cite{seibergparameters}.

\section{The Higgs Sector in Gauge Mediation}

So far, we have adopted the definition of gauge mediation of \cite{meadeseibergshih}, in which, in the limit that the gauge
couplings all tend to zero, there is no supersymmetry breaking in the MSSM sector (for our discussion, we will
again treat $\mu$ as a supersymmetry breaking parameter).  But such a theory cannot be phenomenologically
realistic.  For example, there is necessarily a massless fermion from the Higgs sector, which is not compatible with measurements
of the $Z$ width and other constraints, a relatively light chargino incompatible with experimental
bounds,  and a very light pseudoscalar.  The latter arises
because the theory, in this limit, has an exact symmetry broken only by anomalies, under which the Higgs fields transform by the
same phase.  The Higgs expectation values break this {\it Peccei-Quinn} symmetry, giving rise to a pseudogoldstone boson.

In \cite{meadeseibergshih}, this problem was phrased in terms of operators which would couple Higgs to the messenger
sector, breaking this additional symmetry.
Ref. \cite{sz} provides a systematic analysis of this problem.  If we write renormalizable (dimension three or four) couplings,
then there are two types of couplings between the Higgs fields and messengers:  couplings linear in the Higgs, and those quadratic.
Linear couplings have to involve operators in the messenger sector with weak isospin 1/2; quadratic couplings can involve
singlet or triplet operators.  The problem can, again, be organized in terms of correlation functions in the messenger sector.
However, some of the issues can be illustrated by considering a simple model of elementary messengers coupled to the Higgs:
\beq
W =\lambda_u H_u \ell X + \lambda_d H_d \ell X
\eeq
Here the $X$ field is the same as in eqn \ref{simplegm}.  If $\lambda_u$ and $\lambda_d$ are small, we can integrate out the messenger
fields, already at tree level obtaining a $\mu$ and $B_\mu$ term:
\beq
\mu \sim{\lambda_u \lambda_d} x ~~~~~ B_{\mu} = {\lambda_u \lambda_d} {F \over M}.
\eeq
Now there are several issues.  If we want $\mu$ and $B_\mu$ to be of order the weak scale, $\lambda_u$ and $\lambda_d$ should be small.  From the
expression for $\mu$, we need $\lambda^2$ of order a loop factor.  But this means $\mu^2$ is two loop order, while $B_\mu$ is
one loop order.  This problematic hierarchy is typical of attempts to understand $\mu$ in gauge mediation.   One out, noted in \cite{sz},
is to add additional fields, with a symmetry which suppresses $B_\mu$.  These models also have a potential problem with the $D$ term
for hypercharge, but if $\lambda^2$ is of order a loop factor, then $\langle D_Y \rangle$ is of two loop order, which is small enough.

A number of alternative mechanisms to generate $\mu$ and $B_\mu$ have been proposed.  One class of models, which generate the
correct hierarchy naturally, involves a coupling of a singlet, $S$, to $H_U H_D$.  The singlet also couples to messengers, in such a
way that at two loop order one generates an expectation value for $S$, leading, in turn, to a $\mu$ and $B_\mu$ term, with
$\mu^2 \sim B_\mu$\cite{giudicemu}.

The analysis of \cite{sz} allows a systematic classification of these possibilities, and construction of numerous models.  One can debate
whether these complications make gauge mediation more or less plausible.

\section{Microscopic Models of Supersymmetry Breaking}

Simply introducing soft breaking parameters has two limitations:
\begin{enumerate}
\item  The theory is not complete in the ultraviolet; this is signalled by the logarithmic divergences we observed in soft breaking parameters
\item  Related to the first point, the soft breakings are all independent parameters.
\item  There is no explanation of the large hierarchy.
\end{enumerate}
The gauge mediated models we have introduced to this point represent a significant improvement in that the number
of low energy parameters is greatly reduced, in a manner consistent
with a broad array of experimental constraints.  From the point of view of what we might realistically hope to see at the LHC,
this is perhaps enough.   but  we have not provided any explanation for the
parameters $x$ and $F$ which determine the low energy spectrum.  In the next section, we will provide examples of weakly
coupled, microscopic models
in which these parameters are calculable.  This could be particularly exciting if some
compelling model or dynamical mechanism lead to specific predictions for low
energy phenomena, or if some more direct relic of this dynamics might
be  observable at lower energies.  It may also be of
interest for trying to connect low energy
supersymmetry to some underlying structure (say string theory).

These models will still not address the question of the origin of the hierarchy; we will take
up that question when we discuss dynamical supersymmetry breaking in section \ref{dsb}.
One of the original motivations for considering supersymmetry as a solution of the hierarchy problem is its
susceptibility to {\it dynamical} breaking\cite{wittendsb}.  As we will explain, in (virtually) all situations where
supersymmetry is unbroken at tree level, it is unbroken to all orders of perturbation theory.  But non-perturbative
effects {\it can} break the symmetry.  So one can envisage that the scale of supersymmetry breaking is given
by a formula along the lines:
\beq
m_{susy} = M_{gut} e^{- {8 \pi^2 \over c g^2(M_{gut})}}
\eeq
where $g$ is some gauge coupling and $c$ is an order one constant.  We will see examples of this
sort of phenomenon in section \ref{dsb}.  The relevant dynamical effects can be weak coupling,
semiclassical phenomena (instantons), or strong coupling dynamics.  But we will start by considering models where supersymmetry
is broken at weak coupling, already at tree level.

\section{Weakly Coupled Models of Supersymmetry Breaking}

Before writing down models, it is worth considering some general issues.  We will focus here on models of global supersymmetry,
i.e. we will ignore gravity.  Then the condition for supersymmetry breaking is that one not be able to solve the equations:
\beq
{\partial W \over \partial \phi_i} \ne 0~ \forall~  i.
\eeq
These are {\it holomorphic} equations, i.e. they depend on the $\phi_i$'s, and not there complex conjugates.
It is natural to ask:  for such equations, how is it possible that one might not be able to find solutions?  Nelson and Seiberg
first posed this question in a sharp way, and provided an answer\cite{nelsonseiberg}.  In order that supersymmetry
be broken, it is necessary that some of the equations be inconsistent.  If the superpotential is generic, i.e. if one writes down all possible
terms, and doesn't insist on special relationships among parameters, this does not happen. If there are $N$ fields, one simply obtains
$N$ independent complex equations for $N$ unknowns, and these invariably have solutions.   But if the theory possesses an
$R$ symmetry, then it is possible to obtain inconsistent equations.  Such equations can arise if one has two fields, for example,
$X_1$ and $X_2$, which appear only linearly in the superpotential, i.e.
$$W = X_1 f_1(\phi) + X_2 f_2(\phi)$$
where $\phi$ denotes another field.  Then unless the zeros of $f_1$ coincide with those of $f_2$, the equations
\beq
{\partial W \over \partial X_1} = 0~{\rm and}~ {\partial W \over  \partial X_2} =0
\eeq
are incompatible.

Without a symmetry, though, there is no reason why there shouldn't be terms in the superpotential involving $X_1^2$, $X_1 X_2$, etc.
But an $R$ symmetry can account for a structure like that above.
This is, loosely speaking, the content of the theorem of Nelson and Seiberg, that $R$ symmetries are required, for models with generic
superpotentials, in order to obtain supersymmetry breaking\cite{nelsonseiberg}.
Recall that under an $R$ symmetry, the superpotential carries $R$ charge two.   Then
if one has some number of fields, $X_I$, with $R$ charge two, and the same number with charge $0$, $\phi_i$, one obtains {\it exactly} the structure
we have described above.

Note that there are other possibilities.  For example, if $I=1,\dots,N; i =1,\dots , n$, then if $n \ge N$, one can, generically, satisfy the
equations ${\partial W/\partial X_I} =0$; the $\partial W / \partial \phi_i$ equations are satisfied by simply
setting the $X$'s to zero.  In general, one has an $n-N$ dimensional space of continuous solutions to these equations;
this ``moduli space" survives quantum corrections, at least in perturbtion
theory,  due to supersymmetry (we will discuss
this further shortly).  If  $N>n$, supersymmetry is generically broken, as there are more
equations (for the different $I$'s) than unknowns ($\phi_i$'s).  Classically, there will be an $N-n$ dimensional moduli space of vacua.  Quantum
mechanically, this degeneracy will be lifted; these flat directions are thus referred to as ``pseudomoduli spaces", and
the corresponding fields as pseudomoduli.

\subsection{The O'Raifeartaigh Models}

Models with only chiral fields which implement these ideas are known as O'Raifeartaigh Models.
The simplest such model has three fields, $X_1,X_2,\phi$, and superpotential:
\beq
W = \lambda X_1 (\phi^2 - \mu^2) + mX_2 \phi.
\label{ormodel}
\eeq
This models possess an $R$ symmetry under which $X_i$ have $R$ charge $2$; it is the most general
model consistent with symmetries if there is also a discrete symmetry under which both $\phi$ and $X_2$ are
odd, while $X_1$ is even.
In this model, one has more $X$ type fields than $\phi$ type fields ($N>n$), so supersymmetry is broken.  The equations
\beq
{\partial W \over \partial X_1}=0 ~~~~{\partial W \over \partial X_2} =0
\eeq
are incompatible.  The equation
\beq
{\partial W \over \partial \phi} = 2 \lambda X_1 + m X_2 =0
\label{pseudomoduli}
\eeq
just gives a condition on the $X_i$, which defines the moduli space.

To determine the vacuum value of $\phi$, we need to minimize the potential
\beq
V_{\phi} = \left \vert {\partial W \over \partial X_1} \right \vert^2 + \left \vert {\partial W \over \partial X_2} \right \vert^2
\eeq
$$~~~~ = \vert \lambda \vert^2 \vert \phi^2 - \mu^2 \vert^2
+ m^2 \vert A \vert^2.
$$
If $m^2 > \lambda^2 \mu^2$, than $\phi =0$ at the minimum, and
\beq
\langle V_\phi \rangle = \langle V \rangle = \vert \lambda^2 \mu^4\vert.
\eeq
The order parameter for supersymmetry breaking is
\beq
F_{X_1} = -\lambda \mu^2 ~~~(F_{X_2} =0).
\eeq
$X_1$ is undetermined; $X_2 =0$; this is consistent with the condition for the pseudomoduli of eqn. \ref{pseudomoduli}.
If $m^2 < \lambda^2 \mu^2$, one can readily find the (non-zero) expectation value of $\phi$.

The spectrum reflects the breaking of the supersymmetry.  In the vacua with $\phi=0$, there is a massless chiral multiplet,
$X_1$, and a multiplet with the following features:
\begin{enumerate}
\item   If $X_1 =0$, there is a Dirac fermion (built from the fermionic components of $\phi$ and $X_2$), with mass $m$.
\item   The complex scalars, $\phi$ and $X_2$ (using the same notation here for superfields and their scalar components) have
a potential whose quadratic terms are of the form:
\beq
V_{scalars} = \vert m^2 \vert (\vert \phi \vert^2 + \vert X_2 \vert^2)
\eeq
$$~~~~ + \lambda F_{X_1} (\phi^2 + \phi^{*2})$$
so the scalars $X_2$ are degenerate with the fermions, but the real and imaginary parts of $\phi^2$ are split by
$\lambda F_{X_1}$.  Note that the ``supertrace",
\beq
\sum (-1)^F m_\alpha^2 =0.
\eeq
Here $(-1)^F$ is $1$ for bosons, $-1$ for fermions.
This is a general result, which is easy to prove, for the spectrum of renormalizable theories at tree level.
\end{enumerate}

\vskip .2cm
\noindent
{\bf Exercise:}  By writing the fermion mass matrix for a general theory, and the boson mass matrix, verify the
vanishing of the supertrace.  The proof can be found in \cite{dinebook}, pp. 158-160.
\vskip .2cm

For non-zero $X_1$, these basic features still hold.  The spectrum is particularly simple if $X_1$ is large; then the spectrum
is approximately supersymmetric, with $\phi$ possessing mass $\lambda X_2$, and real and imaginary parts split, in mass-squared,
by $\lambda F_{X_1}$.  The $X_2$ fields have mass of order $m^2/{\lambda X_1}$, and with small splittings.

\vskip .2cm
\noindent
{\bf Exercise:}  Work out the spectrum of the model in the more general vacua for which $X_1 \ne 0$.
\vskip .2cm

\subsection{The Potential on the Pseudomoduli Space}

The lifting of the vacuum degeneracy in this model can be understood very simply.
Classically, the potential is zero.  Quantum mechanically, the vacuum energy receives corrections from the zero point fluctuations
of the bosons, and the Dirac sea of the fermions.  The masses of the fields, as we have just
seen, depend of the value of the pseudmodulus, $X_1$.
As a result, the quantum mechanical contribution to the vacuum energy is a function of $X_1$, i.e. there is a potential for $X_1$.
We can write these two contributions to $V(X_1)$ as:
\beq
V(X_1) = \sum (-1)^F \int {d^3 k \over (2 \pi)^3} {1 \over 2} \sqrt{k^2 + m^2(X_1)}.
\eeq
The individual terms in this expression are divergent, but, due to supersymmetry, there are significant cancelations (these expressions, needless
to say, are quite ill defined; the statement that there are cancelations presupposes a regulator which preserves the symmetry).
The most severe divergence is the mass-independent, quartic divergence, familiar from introductory field theory texts.
This divergence has opposite sign for fermions and bosons, and so cancels due to the equal number of fermionic and bosonic
states.  The subleading, quadratic divergence is:
\beq
\sum (-1)^F \int {d^3 k \over (2 \pi)^3} {1 \over 4} {m^2 \over k}.
\eeq
which cancels due to the supertrace theorem.  Finally, there is a logarithmic divergence, which survives:
\beq
-\sum (-1)^F \int {d^3 k \over (2 \pi)^3} {1 \over 8} {m^4 \over k^3}.
\eeq
The result  is quadratic in $F_{X_1}$; this divergence represents a renormalization of the $X_1$ kinetic term,
$\int d^4 \theta X_1^\dagger X_1$.

We can quickly do the calculation for large $X_1$.   Then the masses are all approximately
equal to $\vert \lambda X_1 \vert$, so we can approximate
the log by $\log \vert X_1 \vert$; $(-1)^F m^4 = 4 \vert \lambda^2 F_x^2\vert$.  So
\beq
V(X_1) \approx   4  {\vert \lambda^2 F_{X_1}^2\vert  \over 4 \pi^2} \log (\vert \lambda^2 X_1 \vert ).
\eeq
This potential grows for large $X_1$.  For smaller $X_1$, it is straightforward to check that the minimum of the potential is at the origin.
This means, in particular, that the $R$ symmetry is not spontaneously broken.  When we come to model-building, this will be
important.

Shih has shown that this is completely general; in models in which all fields have $R$ charge $0$ or $2$, the $R$ symmetry
is always unbroken\cite{shihrsymmetries}.  Models with different $R$ charge assignments can break the symmetry.  The simplest
such model has fields with $R$ charge $2$, $1$, $-1$ and $3$, which we will denote
by  $X$, $\phi_1$, $\phi_{-1}$, $\phi_3$:
\beq
W = - \mu^2 X + \lambda X \phi_{-1} \phi_1 + m_1 \phi_1 \phi_1 + m_2 \phi_{-1} \phi_3.
\label{rbreakingmodel}
\eeq
This model, for a range of parameters, has an $R$-symmetry breaking minimum.

\subsubsection{Complete Models}

With our results above, we can now demonstrate the existence of microscopic models of
models of gauge mediation.  We can simply couple the field $X$ of
the model of eqn. \ref{rbreakingmodel}  to messenger fields,
as in minimal gauge mediation.  In this model, as it stands, CP is unbroken and
the $R$ symmetry is broken, so we can obtain gaugino masses without large phases.
However, it is necessary to introduce additional fields and couplings in order to
generate a $\mu$ term, and this can lead to additional phases and CP violation.

\subsection{Metastable Supersymmetry Breaking}

It is unlikely that any fundamental theory exhibits continuous symmetries; it is a theorem in string theory that there are no global symmetries\cite{banksdixon}.
At best, then, the $R$ symmetry of models like the O'Raifeartaigh models will be approximate.  In the case of the model of eqn. \ref{ormodel}, for example,
the continuous $R$ symmetry might be a consequence of a discrete $R$ symmetry under which
\beq
X_1 \rightarrow e^{2 \pi i \over N} X_1~~X_2 \rightarrow e^{2 \pi i \over N} X_2
\eeq
$$~~~~\theta \rightarrow e^{\pi i \over N} \theta~~~W \rightarrow e^{2 \pi i \over N} W
$$
and $\phi$ is invariant
(this is, again, an $R$ symmetry because the superpotential transforms, and so do the supercharges).
Along with the $X_2 \rightarrow -X_2, \phi \rightarrow -\phi$ symmetry, this accounts for the structure
of the lagrangian, at the renormalizable level.  But couplings like ${1 \over M^{N-2}} X_1^{N+1}$ are allowed by
the symmetry.  As a result, the equation
\beq
{\partial W \over \partial X_1} =0
\label{highdimension}
\eeq
has a solution, with $X_1$ large ($X_1^{N} \sim M^{N-2} \mu^2$).

What of the non-supersymmetric vacuum?  Classically, the high dimension coupling of eqn. \ref{highdimension} destabilizes this state; more precisely it gives
rise to a classical potential on the original moduli space, with a minimum at the supersymmetric point.  However, the quantum mechanical corrections we have
evaluated above render this state a {\it local} minimum of the potential.
The non-supersymmetric vacuum is metastable; it's lifetime is exponentially long, where the exponent scales as a power of $M/\mu$.

Indeed, the Nelson-Seiberg theorem, which requires a continuous R symmetry, and the fact that we don't expect continuous global symmetries in sensible fundamental
theories, suggests that this behavior should be generic.  We will consider this issue again when we discuss models of dynamical supersymmetry breaking.

\section{Dynamical Supersymmetry Breaking}
\label{dsb}

So far, the parameters $\mu$ and $m_i$ in our various models were introduced by
hand.  If they are to be connected to the hierarchy problem, they must be hierarchically small.
Supersymmetry is prone to generating small numbers\cite{wittendsb}.
Most dramatically, if supersymmetry is unbroken at tree level, it is typically unbroken to all
orders of perturbation theory.   This follows from a set of {\it non-renormalization theorems.}
Originally, these theorems were understood by studying the structure of perturbation theory
in supersymmetric theories\cite{supergraphs}; Seiberg explained how to understand
these theorems in a much more conceptual way.  This understanding indicates that the theorems (as originally speculated
by Witten and demonstrated by \cite{ads})  do not
extend beyond perturbation theory, opening the possibility of generating a large hierarchy, of order
$e^{-c/g^2}$.

\subsection{Non-Renormalization Theorems}

Seiberg\cite{seibergnr}, in a program that has had far reaching implications, realized that these theorems could
be understood far more simply.  Moreover, Seiberg's proof indicates clearly when non-perturbative
effects can violate the theorems.  His ingenious suggestion was to consider the couplings in the superpotential,
and the gauge couplings, as expectation values of chiral fields.  These fields must appear holomorphically
in the superpotential and gauge coupling functions, and this greatly restricts the coupling dependence of
these quantities.

To illustrate, consider a simple Wess-Zumino model:
\beq
W = {1 \over 2} m \phi^2 + {1 \over 3} \lambda \phi^3.
\eeq
For $\lambda =0$, this model possesses an R symmetry, under which $\phi$ has $R$ charge $1$.
So we can think of $\lambda$ as a chiral field with $R$ charge $-1$.  Since the
superpotential is holomorphic, the only allowed terms, polynomial in the $\phi$'s, have the form
\beq
\Delta W = \sum_n \lambda^n \phi^{n+3}.
\eeq
This is precisely the $\lambda$ dependence of tree diagrams with $n+3$ external legs; we have predicted that there
are no loop corrections to the superpotential.

Note that there is no corresponding argument for the Kahler potential, and it is easy to check that the Kahler potential is already
renormalized at one loop.  As a result, physical masses and couplings {\it are} corrected in this model.  But the non-renormalization
theorems have, as we will see, profound significance.

For gauge theories, the results are in many ways even more dramatic and surprising.  Again, the coupling can be represented as a complex
field:
\beq
{\cal L} =-{1\over 4} \int d^2 \theta \tau W_\alpha^2
\eeq
where $\tau = {1 \over g^2} + i {\theta \over 8 \pi^2}$
and $\theta$ is the usual CP-violating parameter of the gauge theory.  Perturbation theory is insensitive to $\theta$ (for an introduction to the CP
violating parameter, $\theta$, see, for example, \cite{dinebook}, chapter 5) so, order by
order, the Wilsonian effective action is
\beq
\tau \rightarrow \tau + i \epsilon
\eeq
Apart from ${\cal L}$, the only combination of $\tau$ and $W_\alpha^2$ which is invariant under the shift symmetry is $\int d^2 \theta W_\alpha^2$,
which is precisely the structure of the one loop correction.  So we seem to establish that there is at most a one loop correction to the gauge coupling,
and that {\it there are no loop corrections to the superpotential}.  The gauge coupling result is puzzling, since it is well-known that there is a two
loop correction to the beta-function in supersymmetric gauge theories.  This issue has a resolution, due to Shifman and Vainshtein\cite{sv}; I do not have time to
explain it fully here (see, for example, \cite{dinebook}, pp. 501-503), but heuristically, the issue can be understood by realizing that the {\it cutoff} of the theory, itself,
is not, in general, a holomorphic function of the coupling (equivalently, the question is one of renormalization scheme).

It is well-known that the shift symmetry of perturbation theory is anomalous, and is broken beyond perturbation theory (for an introduction, see \cite{dinebook}, chapter 5).
In an SU(N) gauge theory, for example, instantons generate an expectation value for:
\beq
\langle (\lambda \lambda)^N \rangle  \propto e^{-{8 \pi^2 \over g^2} + i \theta} = e^{-8 \pi^2 \tau}.
\eeq
This expectation value leaves over a discrete $Z_N$ symmetry.
It has long been believed -- and using the sorts of holomorphy arguments developed by Seiberg, shown -- that gluinos condense
in this theory:
\beq
\langle W_\alpha^2 \rangle = \langle \lambda \lambda \rangle = \Lambda^3 e^{i \theta/N} \propto e^{-3 \tau/b_0}.
\eeq

One can think of this as a constant superpotential, so it represents a breakdown of the non-renormalization theorems.
By itself, this is not so interesting.  In global supersymmetry, physics is not sensitive to a constant $W$ (though in local supersymmetry,
if one started in flat space, one would now have a theory with unbroken supersymmetry in anti-De Sitter space).  But now couple
the gauge theory to a singlet, $S$, with no other couplings:
\beq
{\cal L} = (\tau + {S \over M} ) W_\alpha^2.
\eeq
Then
\beq
W_{eff}(S)  \propto e^{-\tau \over 3 b_0}e^{-{1 \over 3 b_0}{S \over M} }
 \eeq
 So a classical moduli space ($S$), has been lifted;
we have breakdown of the non-renormalization theorems and dynamical supersymmetry breaking through non-perturbative effects.

\subsection{Retrofitting of the O'Raifeartaigh Model}

Of course, this is not quite what we are looking for.  The potential for $S$ simply falls to zero at large $S$; if it has a stationary point, it lies in a region
where the effective coupling is large and we cannot calculate.  Moreover, such a stationary point
will not be metastable, as there is no small parameter to suppress the decay rate.

But we can use our experience with O'Raifeartaigh models to build a theory with metastable supersymmetry breaking.
Introduce singlet fields, $X,Y$ and $\phi$, and take for the superpotential and gauge couplings of the model\cite{dfs}:
\beq
{\cal L} = {X \over M} W_\alpha^2 + X \phi^2 + m Y \phi.
\eeq
At energies below the scale of the strong gauge group, $\Lambda$, we can integrate out the gauge interactions
leaving the effective superpotential:
\beq
W(X,A,Y) = X \phi^2 + {\Lambda^3} e^{-{1 \over N} {X \over M}} + mY \phi.
\label{retrofitted}
\eeq
Near $X=0$, this is like the OR model, and one loop corrections will again generate a local minimum
of the potential.  But at very large $X$ ($X$ comparable to $M$), the potential falls away to zero.
A model of this type is said to be {\it retrofitted}.

We would like to account for the mass scale $m$ dynamically as well; we might wish, for example,
that $\mu^2_{eff} = {\Lambda^3 \over M} \sim m^2$.  We will discuss how this can naturally
be achieved in the next subsection, where we discuss the role of symmetries
in such models.

\subsubsection{Symmetries of the Retrofitted Model}

In the case of metastable OR models, we saw that discrete symmetries could account for the approximate
$R$ symmetry of the model.  In the present case, discrete symmetries can also play this role, though
they look somewhat different than those we encountered earlier, since $W_\alpha^2$ necessarily transforms
like the superpotential.  As one possibility, consider:
\beq
W_\alpha \rightarrow e^{2 \pi i \over N} W_\alpha~
X \rightarrow X ~ \phi \rightarrow e^{2 \pi i \over N} \phi~ Y \rightarrow e^{2 \pi i \over N} Y
\eeq
It is necessary to forbid $Y^2$ and $Y^2 X$; this can be achieved by
suitable additional ordinary symmetries (e.g. a $Z_4$, $Y \rightarrow i Y$, $\phi \rightarrow -i \phi$).
The continuous  $R$ symmetry of the low energy theory is then a
consequence of these discrete symmetries, and the restriction
to low order dimension terms in the effective action.

We can account for the scale $m$ of the
low energy theory dynamically
through a structure such as\cite{dinemason}:
\beq
\Delta {\cal L} = \int d^2 \theta \left ({X W_\alpha^2 W_\beta^2 \over M^4} + X \phi^2 + {W_\alpha^2 \over M^2} Y \phi \right ).
\label{wfourth}
\eeq
This model achieves  $m^2 \sim \mu^2$.
Starting with models such as this, we can again couple to messengers and build models of gauge mediated supersymmetry breaking.

\vskip .2cm \noindent
{\bf Exercise:}  Show that the superpotential of eqn. \ref{wfourth} is the most general consistent with symmetries, through
cubic order in the fields.

\vskip .2cm

\section{Supersymmetry Dynamics}

The retrofitted construction is, in some sense, almost too easy.  One could conceive of structures such as this emerging
from a more fundamental theory, but it is worth enlarging our perspective on dynamical supersymmetry breaking.

\subsection{Supersymmetric QCD}

Let's step back and think more about supersymmetric dynamics.
We first consider {\it Supersymmetric QCD} with $N_f$ flavors, which we will define to be a supersymmetric theory with gauge group $SU(N)$
and $N_f$  quarks in the $N$ and $N_f$ in the $\bar N$ representations, $Q_f, \bar Q_{\bar f}$.  Consider, first, the theory with massless quarks.
The model has a global symmetry $SU(N_f) \times SU(N_f) \times U(1)_B \times U(1)_R$.  Here we are listing only symmetries
free of anomalies.  $Q$ and $\bar Q$ transform as
\beq
Q:~(N_f,1,{1 \over N}, {N_f - N \over N_f})
\eeq
$$~~~~~~~ \bar Q:~(1,N_f,-{1 \over N}, {N_f - N \over N_f}).
$$
Let's check the cancelation of anomalies.  We are concerned about
triangle diagrams with the symmetry current at one vertex, and $SU(N)$ gauge bosons
at the other two vertices.    For the $SU(N_f)$ symmetry, the absence of anomalies is automatic (resulting from the
tracelessness of the $SU(N_f)$ generators; for $U(1)_B$ it follows immediately from the opposite baryon numbers of $Q$ and
$\bar Q$.  For the $R$ symmetry, note that the $R$ charge of the gluino is $+1$, so the gluino makes a contribution
to the anomaly proportional to $N$ (the Casimir of the adjoint representation); the $R$ charges of the (fermionic) quarks and antiquarks,
$\psi_Q$ and $\psi_{\bar Q}$ are ${N_f - N \over N_f} -1 = -{N \over N_f}$.
So, as the Casimir of the fundamental is $1/2$, and there are $2 N_f$ fields of the same $R$ charge contributing to the anomaly,
we obtain cancelation.

It is important to understand the structure of the massless theory.  Classically, there
is a large moduli space of SUSY vacua.  The potential arises simply from the
$D^2$ terms of the gauge fields; it is enough to ensure that these vanish.   Up to gauge and flavor transformations,
for $N_f < N$,
\beq
Q = \left ( \matrix{v_1 &0 & 0 &\dots & 0 & \dots  \cr 0 & v_2 & 0 & \dots & 0 & \dots \cr & & & & &\dots \cr 0 & 0 & 0 & \dots & v_{N_f} & \dots  } \right )
\eeq
and $Q = \bar Q$.

\vskip  .2cm
\noindent {\bf Exercise:}
Verify the statement above.  This can be done in a variety of ways.  First, note that
the vanishing of the $D$ terms with $Q,\bar Q$ as above is automatic.  It follows from the ``messenger parity" symmetry we have described above, under
which the $D$ terms are odd.  It also follows from the explicit form of the $D$ terms.  These can be written as $SU(N)$ matrices (see, for example, \cite{dinebook}, chapter 13)
\beq
D^i_j = Q^{*i} Q_j - \bar Q^i \bar Q^*_j - ~{\rm Trace}
\label{dtermequation}
\eeq
One can show this is the most general solution by first using the gauge and flavor symmetries to bring $Q$ to the form, above, and then arguing that, in virtue of
eqn. \ref{dtermequation}, $\bar Q$ is identical, up to flavor transformations.

\vskip .2cm

In these directions, the gauge symmetry is broken to $SU(N-N_F)$.   There are $N^2 - (N-N_f)^2 = 2 N N_f  - N_f^2$ broken generators.
Each broken generator ``eats" one chiral field.  There are also a set of broken flavor symmetries.   For example, if the $v$'s are all equal, the
unbroken flavor symmetry is $U(1) \times SU(N_f)$, so there are $N_f^2-1 + 1$ Goldstone fields; these also arise from the chiral fields; in this way we
have accounted for all of the $Q,\bar Q$ fields.  We can understand this counting another way, by constructing $N_f^2$ gauge invariant ``meson"
fields:
\beq
M_{f,\bar f} = \bar Q_{\bar f} Q_f.
\eeq

Perturbatively, these directions remain flat.  Non-perturbatively, there is a unique superpotential which one can write which is consistent with the symmetries:
\beq
W = {\Lambda^{3 N + N_f \over N - N_f} \over {\rm det} (M_{f,\bar f})}.
\label{wads}
\eeq

In the case $N_f = N-1$, one can compute the superpotential in a straightforward semiclassical analysis.   This is described in \cite{dinebook}, section 13.6.

\vskip .2cm
\noindent
{\bf Exercise}:  Verify that the superpotential of equation \ref{wads} respects the $SU(N_f)$ and non-anomalous $U(1)_R$ symmetries.
Go one step further.  Introduce the background field $\tau$, and assign it a transformation law, $\tau \rightarrow \tau + i~ C~ \alpha$,
choosing the constant $C$ such that
the $\tau$ transformation cancels the anomaly in the symmetry under which $Q$, $\bar Q$ have R charge zero.   Recalling the dependence of $\Lambda$ on
$\tau$, verify that \ref{wads} respects this symmetry as well.

\vskip .2cm

As for our simplest retrofitted model, the potential associated with $W$, while it lifts the moduli, does not yield a stationary point of the potential in any
regime where the coupling is weak.  Again, it is conceivable that there are stationary points in the strong coupling region, but such would-be states
will have unsuppressed decays, i.e. they are not really states at all.

But there are several interesting questions we can study using these results.  First, we can {\it derive} the fact of gaugino condensation.  We'll exhibit
the result for $SU(2)$.  Start with $N=2$, with one flavor.  Then in the general flat direction, the gauge symmetry is completely broken, and
there is a single light meson, $\Phi = \bar Q Q$; the superpotential
can be systematically computed in the semiclassical approximation.    If we include a small mass
for $\Phi$, we have a superpotential
\beq
W = m \Phi + {\Lambda^5 \over \Phi}.
\eeq
This superpotential has supersymmetric stationary points for
\beq
\Phi = e^{2 \pi i \over N}\left ( {\Lambda^5 \over m} \right )^{1/2} ~~~\langle W \rangle = e^{2 \pi i \over N} \Lambda^{5/3} m^{1/2}
\eeq
where we have written the phase to point to the generalization to the case of $SU(N)$ with $N-1$ flavors.  Now Seiberg points out that
$\langle W \rangle$ is a holomorphic function of $m$ and $\Lambda$, and it must transform properly under the symmetries (with $m$ and
$\Lambda$ treated as background fields).  As a result, this expression is valid for all $m$, and in particular for $m$ large, so that one
is studying, at low energies, a pure $SU(2)$ gauge theory.  In that theory, $\langle W \rangle = \langle \lambda \lambda \rangle$.

\vskip .2cm
\noindent
{\bf Exercise:}  Verify that the dependence on $m$ and $\Lambda$ of $\langle W \rangle$ is uniquely determined by symmetries (see
\cite{dinebook}, p. 208).
\vskip .2cm

\subsection{Varieties of Dynamical Supersymmetry Breaking}

Modifying slightly the theories we have designated as Supersymmetric QCD, we will be able to uncover two types of supersymmetry breaking.
One is stable, in the sense that, at the renormalizable level (more generally, up to operators of some fixed
dimension), there are only supersymmetry-breaking ground states (related to one another by symmetries);
the second is metastable\cite{iss}, much like the theories we have encountered in the weakly coupled and retrofitted cases, where,
even at the renormalizable level,
there are additional supersymmetric states, separated by a large barrier or a sizable distance in field space.

\subsubsection{Stable Supersymmetry Breaking -- the $3-2$ Model}

Models with stable supersymmetry breaking are rare.  They are generally characterized by two features:
\begin{enumerate}
\item  Classically, their potentials have no flat directions (there is not a moduli space of vacua).
\item  They exhibit global symmetries, which are spontaneously broken in the ground state.
\end{enumerate}
If the first condition is {\it not} satisfied, then, as we have seen, there are typically regions in the moduli space where the potential
tends to zero, corresponding to (at least) asymptotic restoration of supersymmetry.  If the second condition is satisfied, there
is some number of Goldstone particles.  In general, as in the example of massless QCD, these particles each lie in a different
chiral multiplet.  The other scalar in the multiplet, like the Goldstone, will have no potential; it is a flat direction, contradicting the
first assumption above.  There is a potential loophole in this argument:  it is logically possible that both fields in the multiplet
are Goldstone particles.  Typically, however, this does not occur.

The simplest example of such a theory, in which it is possible to do systematic calculations, is known
as the $3-2$ model because the gauge group is $SU(3) \times SU(2)$.  Its particle content is like that of
a single generation of the Standard Model, minus the singlet:
\beq
Q:(3,2)~\bar U:(\bar 3,1) ~\bar D:(\bar 3,1)~L = (1,2).
\eeq
There is a unique superpotential allowed by the symmetries, up to field redefinitions:
\beq
W = \lambda Q L \bar U.
\label{32w}
\eeq
Without the superpotential, and assuming that the $SU(2)$ coupling is much smaller than the $SU(3)$ coupling,
this is supersymmetric QCD with $N=3,N_f =2$.
The theory has a set of flat directions, and generates a non-perturbative superpotential.  It is easy to see, however,
that the {\it classical} superpotential of eqn. \ref{32w} already lifts all of the flat directions.  Moreover, the theory
possesses a non-anomalous $R$ symmetry.

\vskip .2cm
\noindent
{\bf Exercise:}
First show that, without loss of generality,
you can take, for the flat direction of the $D$-term:
\beq
Q = \left ( \matrix{a & 0 & 0 \cr 0 & b & 0} \right )~~~~ L = \left ( \matrix{e^{i \phi} \sqrt{\vert a \vert^2 - \vert b \vert^2}\cr 0} \right )
\eeq
$$\bar U = \left ( \matrix{a & 0 & 0} \right )~~\bar D = \left ( \matrix{0 & b & 0} \right ).$$
Now show that for any choice of $\bar U$, one cannot satisfy all of the ${\partial W \over \partial \phi_i}$'s to zero, unless both $a$ and
$b$ vanish.  Finally, check that the model possesses a non-anomalous $R$ symmetry which is spontaneously broken by a non-vanishing
$a$ or $b$.

\vskip .2cm

For small $\lambda$, the effective superpotential is:
\beq
W_{eff} = {\Lambda^6 \over Q Q \bar U \bar D} + \lambda Q L \bar U.
\eeq
Careful study of the resulting potential exhibits a supersymmetry-breaking minimum.

One can ask:  what happens in this model if the $SU(2)$ coupling is much greater than the $SU(3)$ coupling, so that the $SU(2)$ gauge group becomes strong first.
In this limit, the theory looks like $QCD$ with $N=2, ~N_f=2$.  In this theory, there is no non-perturbative superpotential:  there
exists an exact moduli
space, even quantum mechanically.  However, as Seiberg showed, in such theories, the moduli space is modified quantum mechanically.
In effect, $QL$ is non-zero everywhere on the moduli space, generating an $F$ term for $\bar U$ through eqn. \ref{32w}\cite{intriligatorthomas} ( for a pedagogical
discussion of the 3-2 model, including this issue, see \cite{dinebook}, sections 14.1, 16.3.1).

Finally, what if we give up the requirement of renormalizability?  Adding higher dimension operators (e.g. $(Q \bar U L)^2$), with
coefficients scaled by a large mass, $M$,  we will lose the
continuous $R$ symmetry as an exact symmetry, and there will be supersymmetric minima.  These minima, however, will be far away, and separated
by a large barrier (with barrier height scaled by $M$) from the non-supersymmetric minimum near the origin.  The metastable state near the origin will be extremely stable.

It is not easy to find theories which satisfy the conditions for stable supersymmetry breaking, and those which exist pose challenges for model building.
To illustrate the issues, we can consider a class of models with gauge group $SU(N)$, and an antisymmetric tensor, $A_{ij}$, as well as $N-4$ antifundamentals, $\bar F$.
The simplest of these models has $N=5$, and a single $\bar 5$.  It is easy to check, using the matrix technique developed above, that there are no
flat directions of the $D$ terms.   There is a non-anomalous $R$ symmetry; one can give arguments that, in this strongly coupled theory, the symmetry
is spontaneously broken\cite{adssu5}.  These features extend to the model with general $N$, when we
include the most general superpotential
\beq
W = \lambda_{ab} \bar F^a_i \bar F^b_j A^{ij}.
\eeq
So we might adopt the following strategy for model building.  Take $N_f$ (and hence $N$) and choose $\lambda$ appropriately so that the model
has a large flavor group.  For example, for $N=14$, the flavor group can include $SU(5)$; gauge a subgroup of the flavor group, and identify this subgroup with the
Standard Model gauge group.  The difficulty is that the QCD and other gauge couplings are, in this case, violently non-asymptotically free.  Unification is lost;
 indeed, some enlargement of the Standard Model group is required only a few decades above the susy-breaking scale.  Most model building with
 stable supersymmetry breaking invokes more complicated structures in order to obtain a vev for a field like $X$, which in turn couples to messengers.
 The constructions are rather baroque\cite{dinenelson}.

\subsubsection{Metastable Supersymmetry Breaking:  The ISS Model}

We have seen that for $N_f < N$, there is a potential generated non-perturbatively on the classical
moduli space.  For $N_f \ge N$, this is not the case; there is always an exact moduli space.  But the dynamics
on this moduli space, particularly in the region of strong coupling, is quite intricate, with a strong dependence
on the values of $N$ and $N_f$\cite{seibergpapers} (for short, pedagogical introductions, see \cite{peskintasi} and
\cite{dinebook}, chapter 16).

One interesting range is $ N+1 < N_f < 3/2 N$.  Here, the theory is dual to a theory
with gauge group $SU(N_f -N)$, with $N_f$ flavors of quarks, $q_f$ and $\bar q_f$, and a set of mesons,
$M_{f,\bar f}$; the effective lagrangian of this theory possesses a superpotential:
\beq
W= \bar q M q.
\eeq
This duality is not meant as an exact equivalence, but rather a statement about the infrared behaviors of the two
theories.

We have seen that massive QCD has $N$ supersymmetric vacua, connected with the breaking of the $Z_N$ symmetry
of the theory.  Long ago, Witten, in fact, proved that there were $N$ such vacua, and as a result, these theories
were not viewed as an interesting arena for dynamical supersymmetry breaking\cite{wittenconstraints}.
But Intriligator, Shih and Seiberg\cite{iss} made the following remarkable observation about theories in this range
of $N_f$.  They considered adding a small mass term for the quarks, $m \bar Q Q$ in the ultraviolet theory.
Then, in the infrared, ``magnetic", theory the superpotential is:
\beq
W_{mag} =\bar q M q + {\rm Tr}~ m \Lambda M.
\eeq
This superpotential does not have supersymmetric minima.

\vskip .2cm  \noindent
{\bf Exercise:}  Verify that the additional term transforms correctly under the $SU(N_f)$ and $U(1)_R$ symmetries
of the underlying theory.

\vskip .2cm  \noindent
{\bf Exercise}:  Show that the $D$ term conditions and $F$ term conditions in the magnetic theory cannot be satisfied
simultaneously.  One can first bring $q$ to the following form, using gauge and flavor transformations
\beq
q_f^i = v_i \delta_{if} ~~f < i; ~~{\rm zero~otherwise}
\eeq
Then the vanishing of the $D$ terms allows one to bring $\bar q$ to the same form, up to flavor transformations.  Now it is straightforward
to see that the ${\partial W /\partial M_{f \bar f}}$ conditions cannot be satisfied, for $f > N$.  This type of breaking is called
``rank breaking" by the authors of \cite{iss}.

\vskip .2cm

At the classical level, the magnetic lagrangian gives rise to moduli; $M_{f \bar f}$ is not fixed, for example.  Almost as remarkable as the
fact that supersymmetry is broken is the fact that one can compute the potential for $M, q \bar q$ near the origin, {\it even though the theory
is not weakly coupled.}  The result of this computation is that the minimum of the potential does not break the $R$ symmetry (the expectation
value of $M$ vanishes).  For model building, this is problematic, at least if supersymmetry is to be broken at a low scale.
But this is likely the tip of the iceberg of a large class of strongly coupled models exhibiting metastable supersymmetry breaking.
\footnote{A model building program with a low scale ISS-type model has been developed by Banks and collaborators\cite{pentagon}.}

We should  pause and ask:  where are the $N$ supersymmetric vacua in the magnetic description?  These can be understood in the dual
picture.  We remarked that in the metastable vacuum, the $R$ symmetry is unbroken, but we expect that the supersymmetric vacua should
exhibit broken discrete $R$ symmetries.  So consider giving $M$ a large expectation value.  Then all of the dual quarks are massive,
and the low energy theory is asymptotically free.  Gaugino condensation leads to an additional term in the superpotential for $\Phi$,
and one finds $N$ supersymmetric vacua.

\section{Conclusions}

Physicists have been considering supersymmetry as a major component of electroweak symmetry breaking for nearly 30 years.
The ideas surrounding gauge mediation are nearly as old, and models of dynamical supersymmetry breaking have been known for 25 years.
Yet the last three years have seen significant developments. The appreciation that metastable supersymmetry breaking is likely
to be an important component of supersymmetry breaking has opened up vast possibilities for model building; with this has come the
realization that even for gauge mediation, the possible spectra and phenomenology\cite{linda} is much richer than might have been imagined.

As of this writing, the start of physics from the LHC looks to be about one year away.
It is possible that we are on the brink of resolving the origin of electroweak symmetry breaking and the hierarchy problem;
if so, supersymmetry likely plays some role.  But given that the scale of supersymmetry breaking might reasonably be expected
to be of order $M_Z$, it is troubling that we have seen no direct evidence of supersymmetry already.  I have tried to suggest,
in these lectures, that while this lack of evidence may well be argue against supersymmetry, it may instead be telling us that
underlying supersymmetry breaking is dynamics at some relatively low energy scale, mediated to ordinary fields by gauge interactions.

\noindent
{\Large  {\bf Acknowledgements}}

I am grateful to my collaborators on the projects alluded to here.  In recent years, these include
Linda Carpenter, Jonathan Feng, Guido Festuccia, John Mason, Nathan Seiberg,  Eva Silverstein and Scott Thomas.  This work supported in part by the U.S.
Department of Energy.

\end{document}